\documentclass[nofootinbib,aps,pra,superscriptaddress,reprint]{revtex4-2}
\usepackage{amsmath,amsthm,amsfonts,amssymb}
\usepackage{mathptmx}
\usepackage{graphicx}
\usepackage{bm}
\usepackage{braket} 
\usepackage{fancyhdr} 
\usepackage{soul} 
\usepackage{times} 
\usepackage{color} 
\usepackage[colorlinks,linkcolor=blue,anchorcolor=blue,citecolor=blue,urlcolor=black]{hyperref}

\newcommand{\bigchi}{\makebox{\large\ensuremath{\chi}}}
\DeclareMathAlphabet{\mathcal}{OMS}{cmsy}{m}{n}
\begin{document}
\title{On the Teleportation of  Discrete Variable Qubits Via  Continuous Variable Lossy Channels}

\date{\today}

\author{Mingjian He}
\email{mingjian.he@unsw.edu.au}

\author{Robert Malaney}
\email{r.malaney@unsw.edu.au}
\affiliation{School of Electrical Engineering and Telecommunications,\\
	The University of New South Wales, Sydney, NSW 2052, Australia.}

\author{Ryan Aguinaldo}
\email{ryan.aguinaldo@ngc.com}
\affiliation{Northrop Grumman Corporation, San Diego, CA 92128, USA.}

\begin{abstract}
The Continuous-Variable (CV) quantum state of light allows for the teleportation of a Discrete-Variable (DV) photonic qubit. Such an operation is useful in the realm of hybrid quantum networks.
However, it is known that the teleportation of a DV qubit via a Gaussian CV resource channel is severely limited under channel loss, with a teleportation fidelity beyond the classical limit unattainable for losses exceeding a small threshold of 0.5~dB.
In this work, we present a new non-deterministic teleportation protocol that combines a Gaussian CV resource channel with a modified form of the Bell State Measurement that accommodates a DV-qubit fidelity beyond the classical limit for channel losses up to 20~dB. Beyond this orders of magnitude improvement, we also show how the use of non-Gaussian operations on the CV resource channel can lead to a DV-qubit fidelity approaching unity for any channel loss.		
\end{abstract}

\maketitle

\thispagestyle{fancy}
\pagestyle{fancy}
\renewcommand{\headrulewidth}{0pt}	


\section{Introduction}
Quantum teleportation is the process of transmitting unknown quantum states from one location to another.
At its core,  is the consumption of entanglement shared between the locations.
Teleportation can be classified into three categories depending on whether Discrete-Variable (DV) entanglement \cite{bennett1993teleporting,marshall2014high}, Continuous-Variable (CV) entanglement \cite{vaidman1994teleportation,braunstein1998teleportation}, or a combination (hybrid) of DV and CV entanglement is used as the resource channel for the teleportation \cite{brask2010hybrid,lim2016loss,ulanov2017quantum,podoshvedov2019efficient}.

A non-Gaussian resource channel built on DV entanglement was originally proposed for the teleportation of DV qubits \cite{bennett1993teleporting}.
However, in the context of optical communications such a  teleportation scheme cannot be implemented in a deterministic manner using only linear optics  \cite{lutkenhaus1999bell}. In principle, more sophisticated DV-only schemes, which approach unit fidelity, are available for deterministic DV-qubit teleportation,  albeit with no clear practical route to deployment \cite{lee2015nearly}. In contrast,
it was proposed in \cite{polkinghorne1999continuous} and demonstrated in \cite{takeda2013deterministic}, that the deterministic teleportation of DV qubits is possible with an alternative entanglement resource, the Gaussian Two-Mode Squeezed-Vacuum (TMSV) state, albeit at the cost of non-unit fidelity (except in the unattainable infinite-squeezing limit).

However, recent work shows that the teleportation fidelity of DV qubits via a TMSV resource channel is severely limited under a lossy environment -- fidelity cannot reach the classical limit when the resource channel loss exceeds 11\% (0.5~dB) even with infinite squeezing \cite{lie2019limitations}. 
Such sensitivity to loss greatly limits the potential of DV-qubit teleportation via a TMSV resource - an outcome that inhibits the useful operation of heterogeneous quantum networks in which quantum information will be exchanged between DV-enabled and CV-enabled devices \cite{sychev2018entanglement,guccione2020connecting,do2021satellite}.
This is somewhat disappointing as it is believed that TMSV resource channels between the different types of quantum devices could be easily delivered and utilized as the ``inter-connect channel'' in emerging hybrid quantum networks e.g., \cite{andersen2013high,takeda2015entanglement, 9463774, rr1},
 and play an important role in hybrid quantum information processors - particularly in the realm of optical measurement-based quantum computation \cite{rr2}.

Regarding the teleportation of a DV qubit via a TMSV resource channel, it is therefore interesting to ask: Do improved loss-resistant schemes that retain reasonable success probabilities exist? If in the affirmative, such an outcome could have profound implications for a range of quantum information protocols.
In this work, we indeed answer this question in the affirmative by presenting a new non-deterministic teleportation protocol that combines a TMSV resource channel with a hybrid form of a Bell State Measurement (BSM). Henceforth, we refer to this measurement as the H-BSM, and its use in  teleportation as the H-BSM protocol. As we shall see, the replacement of the CV-BSM with our H-BSM,  leads to dramatic improvements in  DV-qubit teleportation fidelity for a range of lossy-TMSV resource channels, with the classical limit being surpassed even when the resource-channel losses exceed 99\% (20~dB).

Beyond the use of our H-BSM, for comparison, we also study the use of distillation operations on the resource channel via Quantum Scissors (QS) and Photon Catalysis (PC)\footnote{We focus on QS and PC since  both have been shown to function as a noiseless linear amplifier for states with low photon number 
\cite{ralph2009nondeterministic,ulanov2015undoing,zhang2018photon,hu2019entanglement}.},
as a means to improve the traditional CV teleportation scheme, as applied to DV qubits.
Such non-Gaussian operations have been shown to improve the teleportation of CV states (e.g., \cite{cochrane2002teleportation, yang2009entanglement,dell2010realistic,zhang2010distillation,seshadreesan2015non,wang2015continuous,xu2015enhancing,hu2017continuous,bose2021analysis}).
However, it remains unclear whether such distillation operations improve the teleportation of DV qubits.
As we shall see, when applied to the TMSV resource channel, both QS and PC can further improve the teleportation fidelity for DV qubits.
More specifically, we will see that a teleportation fidelity approaching unity is possible with QS or PC.

Our teleportation protocols are non-Gaussian in nature and require the use of number-resolving single-photon detectors. 
However, the implementation of single-photon detection with perfect efficiency is still challenging \cite{pernice2012high,esmaeil2017single,slussarenko2019photonic,mardani2020continuous}.
To determine the performance of our protocols under less challenging settings, we also study their use with non-ideal single-photon detectors, showing that teleportation fidelity outcomes above the classical limit are still achievable under a range of detector-efficiency conditions.

The rest of the paper is organized as follows: In Section~\ref{sec:protocols}, we compare existing teleportation protocols with our new protocol. 
The use of additional non-Gaussian operations in our protocol is studied in Section~\ref{sec:non-Gaussian}.
The impact of single-photon detectors with non-unity efficiency is studied in Section~\ref{sec:ineffdetector}.
Section~\ref{sec:conclusion} concludes our work.

\section{Teleportation of DV Qubits with TMSV States}\label{sec:protocols}
\begin{figure}
	\centering
	\includegraphics[width=.95\linewidth]{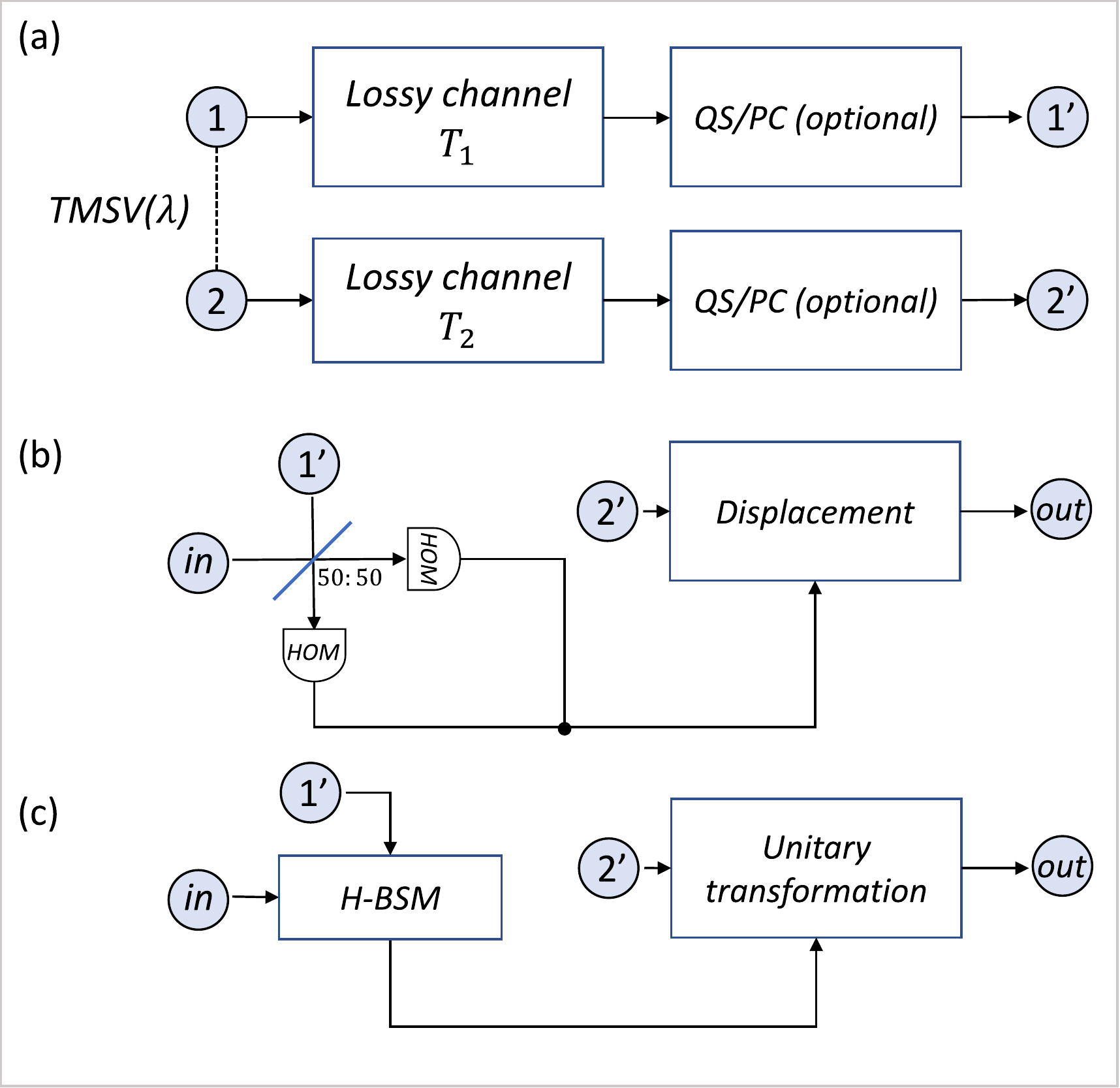}
	\caption{(a) Distribution of a TMSV state with modes $1$ and $2$ over lossy channels with transmissivities $T_1$ and $T_2$.
		The distributed TMSV state can be distilled at the receivers by QS or PC. 		
		(b) The CV-BSM protocol. The distributed TMSV state with modes $1'$ and $2'$ is used as the teleportation resource channel for an input DV qubit. 
		(c) The H-BSM protocol, which also uses the TMSV  teleportation resource channel (HOM: homodyne detector.)}
	\label{fig:diagtphom}
\end{figure}	
First, consider the use of the protocol proposed by \cite{vaidman1994teleportation} in the teleportation of DV qubits expressed in the Fock basis as $\ket{\mathrm{in}}=\cos(\theta/2)\ket{0}+\exp{(i\phi)}\sin(\theta/2)\ket{1}$.\footnote{The term ``DV'' is used to discriminate between the particle-like qubit and the wave-like qubit \cite{morin2014remote}. In this work, we only consider the particle-like qubit that resides in a finite-dimensional space spanned by the absence or presence of a single photon. For conciseness, we will drop the term ``DV'' in the rest of the paper.}
In this protocol, a TMSV state is used as the teleportation resource channel.
An input qubit  (the state being teleported) is coupled with one mode of the resource channel at a 50:50 beam-splitter.
A CV-BSM with homodyne detectors is then performed on the coupled modes.
Based on the measurement result, a displacement operation is applied to the other mode of the resource channel.
In the limit of infinite squeezing, the multi-photon components of the displaced mode (the output mode) vanish and it approaches the input qubit. 
We refer to this protocol as the CV-BSM protocol.

The relation between the input qubit and the output mode for the CV-BSM protocol can be concisely described using the characteristic function formalism.
 Consider a realistic scenario where two locations do not share any entangled state before teleportation, and a TMSV state is first prepared at a middle station (see  Fig.~\ref{fig:diagtphom}a). The characteristic function for the TMSV state can be written as
\begin{equation}
	\begin{aligned}
		\bigchi_{\mathrm{TMSV}}(\xi_1, \xi_2)=&\exp \Big\{ -\frac{1}{2}\Big[\frac{1+\lambda^2}{1-\lambda^2} (|\xi_1|^2 + |\xi_2|^2 )\\ 
		&- \frac{2\lambda}{1-\lambda^2}(\xi_1 \xi_2 + \xi_1^* \xi_2^*) \Big] \Big\},	
	\end{aligned}
	\label{eq:CF_TMSV}
\end{equation}
where $\xi_1$ and $\xi_2$ are complex variables, $\lambda=\tanh{r}$, and $r>0$ is the squeezing parameter for the TMSV state.
The two modes of the TMSV state are then sent through two independent lossy channels characterized by the transmissivities $T_1$ and $T_2$, respectively.
The characteristic function for the state after the channel transmission can be written as \cite{li2011time}
\begin{equation}\label{eq:tmsv_lossy}
	\begin{aligned}
		\bigchi_{\mathrm{TMSV}}'(\xi_1, \xi_2)=&\exp{\left\{-\frac{1}{2}\left[(1 - T_1)|\xi_1|^2+(1 - T_2)|\xi_2|^2\right]\right\}}\\
		&\times\bigchi_{\mathrm{TMSV}}(\sqrt{T_1}\xi_1,\sqrt{T_2}\xi_2).
	\end{aligned}
\end{equation}
As shown in Fig.~\ref{fig:diagtphom}b, the entangled state with the above characteristic function is then used as the teleportation resource channel.
Let $\bigchi_{\mathrm{in}}(\xi)$ and $\bigchi_{\mathrm{out}}(\xi)$ be the characteristic functions of the input qubit and the output mode, respectively.
The latter can then be written as \cite{dell2010realistic}
\begin{equation}\label{eq:cfinandout}
	\bigchi_{\mathrm{out}}(\xi)= \bigchi_{\mathrm{in}}(g \xi) \bigchi_{\mathrm{TMSV}}'(\xi, g  \xi^*),
\end{equation}
where $g$ is the displacement gain factor.
Assume the input qubit is pure.
For this CV-BSM protocol, the teleportation fidelity, which measures the closeness between the characteristic functions of the input qubit and the output mode, can then be written as 
\begin{equation}\label{eq:fidelitycf}
	\mathcal{F} = \frac{1}{\pi} \int d^2 \xi \bigchi_\mathrm{in}(\xi) \bigchi_\mathrm{out}(-\xi).	
\end{equation}

Next, consider our new protocol, the H-BSM protocol of  Fig.~\ref{fig:diagtphom}c. This uses the same teleportation resource channel as before, but with a different strategy on the measurement of the input qubit and the correction of the output mode.
As shown in Fig.~\ref{fig:diagtphom}c, the H-BSM is performed on the input qubit and one mode of the resource channel. Note, our hybrid qubit-mode H-BSM is different from the usual DV-BSM considered for a qubit-qubit system. The different input states to the H-BSM lead to a range of alternate outcomes, in terms of success probabilities, dependent on the setup of the experiment.
Depending on the  results from the H-BSM, a unitary transformation is then applied to the output mode to recover the input qubit.
For a given input qubit $\hat{\rho}_\mathrm{in}=\ket{\mathrm{in}}\bra{\mathrm{in}}$, the output mode for our  H-BSM protocol can be obtained as follows.\footnote{ 
Different from the CV-BSM protocol, the bra-ket formalism is adopted since the characteristic function formalism does not provide a concise description of the relation between the input and output for this new protocol.}

In the Fock basis, the initial TMSV state prepared at the middle station can be written as
\begin{equation}
	\ket{\mathrm{TMSV}}=\sqrt{1-\lambda^2}\sum_{n=0}^{\infty}\lambda^n\ket{nn}_{12}.
\end{equation}
The two lossy channels alter the TMSV state to \cite{sabapathy2011robustness}
\begin{equation}\label{eq:rhotmsvlossy}
	\hat{\rho}_{\mathrm{TMSV}}'=\mathcal{T}_1\circ\mathcal{T}_2(\hat{\rho}_{\mathrm{TMSV}}),
\end{equation}
where $\hat{\rho}_{\mathrm{TMSV}}=\ket{\mathrm{TMSV}}\bra{\mathrm{TMSV}}$, $\circ$ represents the composition of transformations, and the transformation $\mathcal{T}_k$ ($k=1, 2$) is defined as
\begin{equation}\label{eq:lossyopeator}
	\mathcal{T}_k(\hat\rho):=
	\sum_{l=0}^\infty 
	\hat{G}^{(l)}_{k}
	\hat{\rho}
	\hat{G}^{(l)\dagger}_{k},
\end{equation}
with
\begin{equation}
	\hat{G}_{k}^{(l)}=\frac{\hat{a}^l_k\sqrt{T_k}^{\hat{a}_k^\dagger\hat{a}_k}}{\sqrt{l!}}\sqrt{\frac{1-T_k}{T_k}}^l,
\end{equation}
and $\hat{a}_k$ the annihilation operator of mode $k$. For clarity, the two modes of the TMSV state, after passing through the channels, are labeled as $1'$ and $2'$, respectively.
An H-BSM, which detects the four Bell states,
\begin{equation}
\begin{aligned}
    \ket{\Phi^\pm}&=\frac{1}{\sqrt2}\left(\ket{00}\pm\ket{11}\right)\\
    \ket{\Psi^\pm}&=\frac{1}{\sqrt2}\left(\ket{01}\pm\ket{10}\right),
\end{aligned}
\end{equation}
is then performed on the input qubit and mode $1'$.
Depending on the result of the H-BSM, a unitary transformation chosen from 
\begin{equation}
	\begin{aligned}
 		\hat{\sigma}_I&=\ket{0}\bra{0}+\ket{1}\bra{1}\ (\mathrm{for} \ket{\Phi^+})\\
 		\hat{\sigma}_Z&=\ket{0}\bra{0}-\ket{1}\bra{1}\ (\mathrm{for} \ket{\Phi^-})\\
		\hat{\sigma}_X&=\ket{0}\bra{1}+\ket{1}\bra{0}\ (\mathrm{for} \ket{\Psi^+})\\
		\hat{\sigma}_{ZX}&=\ket{0}\bra{1}-\ket{1}\bra{0}\ (\mathrm{for} \ket{\Psi^-})
	\end{aligned}
\end{equation}
is applied to mode $2'$ of the teleportation resource channel.
On the detection of $\ket{\Phi^+}$ for the H-BSM, the output mode of our protocol can be written as
\begin{equation}
	\hat{\rho}_\mathrm{out}=
	\hat{\sigma}_I
	\frac{1}{P_{\Phi^+}}\left(\bra{\Phi^+}\hat{\rho}_\mathrm{in}\otimes\hat{\rho}_{\mathrm{TMSV}}'\ket{\Phi^+}\right)
	\hat{\sigma}_I^\dagger,
\end{equation}
where $P_{\Phi^+}$ is the probability for the detection of $\ket{\Phi^+}$. 
Similar definitions of $P_{\Phi^-}$, $P_{\Psi^+}$, and $P_{\Psi^-}$ are in place for the detection of $\ket{\Phi^-}$, $\ket{\Psi^+}$, and $\ket{\Psi^-}$, respectively.

The measurement basis of our  H-BSM is not complete in the infinite-dimensional space of mode $1'$.
Consequently, the measurement is non-deterministic and its success probability  decreases with increasing mean photon number in mode $1'$.
In this section, we assume the H-BSM that only discriminates the Bell states $\ket{\Psi^\pm}$ is adopted. This is the simplest measurement possible, and one that is readily implemented using current technology.
An implementation of the  H-BSM that discriminates the four Bell states, and one which requires additional non-Gaussian operations,  will be discussed in the next section.
Details on the implementation of  both forms of the H-BSM can be found in Appendix A.

We define the total success probability for the H-BSM protocol as
\begin{equation}
    P_{\mathrm{total}}=P_{\mathrm{BSM}}P_{\mathrm{operation}},
\end{equation}
where 
\begin{equation}
    P_{\mathrm{BSM}}=P_{\Phi^+}+P_{\Phi^-}+P_{\Psi^+}+P_{\Psi^-},    
\end{equation}
is the success probability for the H-BSM, 
and $P_{\mathrm{operation}}$ is the success probability for  additional non-Gaussian operations (beyond photon detection).
Mode $2'$ will be discarded when the protocol fails.
For the H-BSM considered in this section, $P_{\Phi^+}=P_{\Phi^-}=0$, and $P_{\mathrm{total}} = P_{\mathrm{BSM}}$ because no additional operation is adopted.

For a pure qubit as the input (the state being teleported) to our protocol, the teleportation fidelity can be written as 
\begin{equation}\label{eq:fidelityden}
	\mathcal{F} = \mathrm{tr}\left\{\hat{\rho}_\mathrm{in}\hat{\rho}_\mathrm{out}\right\},
\end{equation}
which is equivalent to the fidelity defined in Eq.~(\ref{eq:fidelitycf}). 
Assume the input qubit is uniformly distributed on the Bloch sphere. Then the joint probability density function for the parameters $\theta$ and $\phi$ can be written as $P(\theta,\phi)=(\sin\theta)/4\pi$.
The averaged fidelity, $\bar{\mathcal{F}}$, will be used as the performance metric.
For the H-BSM protocol, $\bar{\mathcal{F}}$ is obtained by first averaging the $\mathcal{F}$ in Eq.~(\ref{eq:fidelityden}) over $\theta$, $\phi$, and the two possible outcomes of the H-BSM.
The averaged fidelity is then normalized by $P_\mathrm{BSM}$.
For the CV-BSM protocol, $\bar{\mathcal{F}}$ is obtained by averaging the $\mathcal{F}$ in Eq.~(\ref{eq:fidelitycf}) over $\theta$ and $\phi$.

First, consider the scenario where the lossy channels are symmetric, i.e., $T_1=T_2=T$. In the following figures,
the channel loss in dB units is determined via $T\mathrm{[dB]}=-10\log_{10}\left(T\right)$, and the
 initial squeezing of the TMSV states in  dB units  is determined via $r\mathrm{[dB]}=10\log_{10}[\exp(2r)]$. In all calculations, the TMSV state is truncated to a finite dimension such that $\mathrm{tr}(\hat{\rho}_{\mathrm{TMSV}})>0.95$ is satisfied.

\begin{figure}
	\centering
	\includegraphics[width=0.95\linewidth]{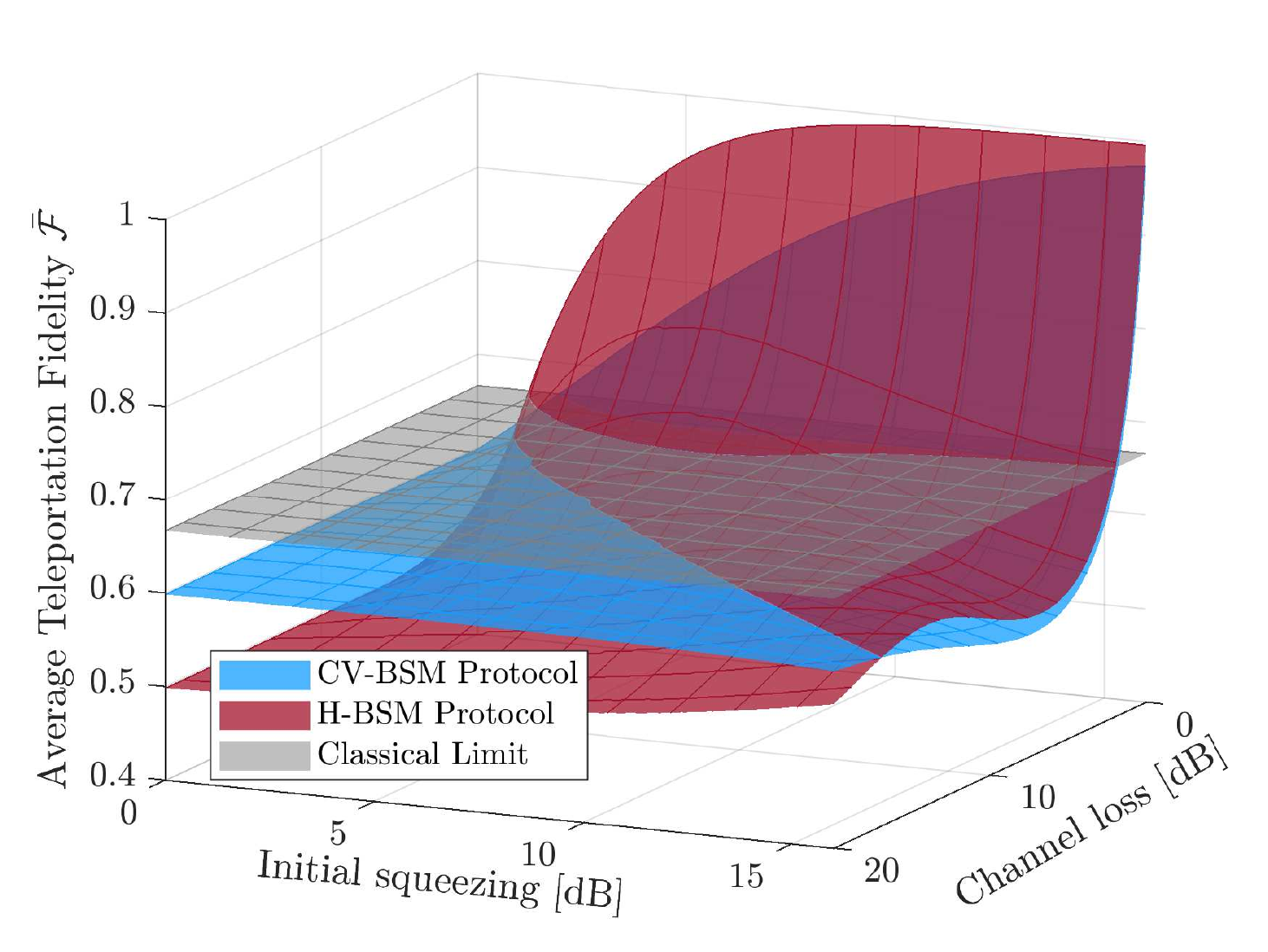}
	
	(a) \vskip 10pt
	
	\includegraphics[width=0.95\linewidth]{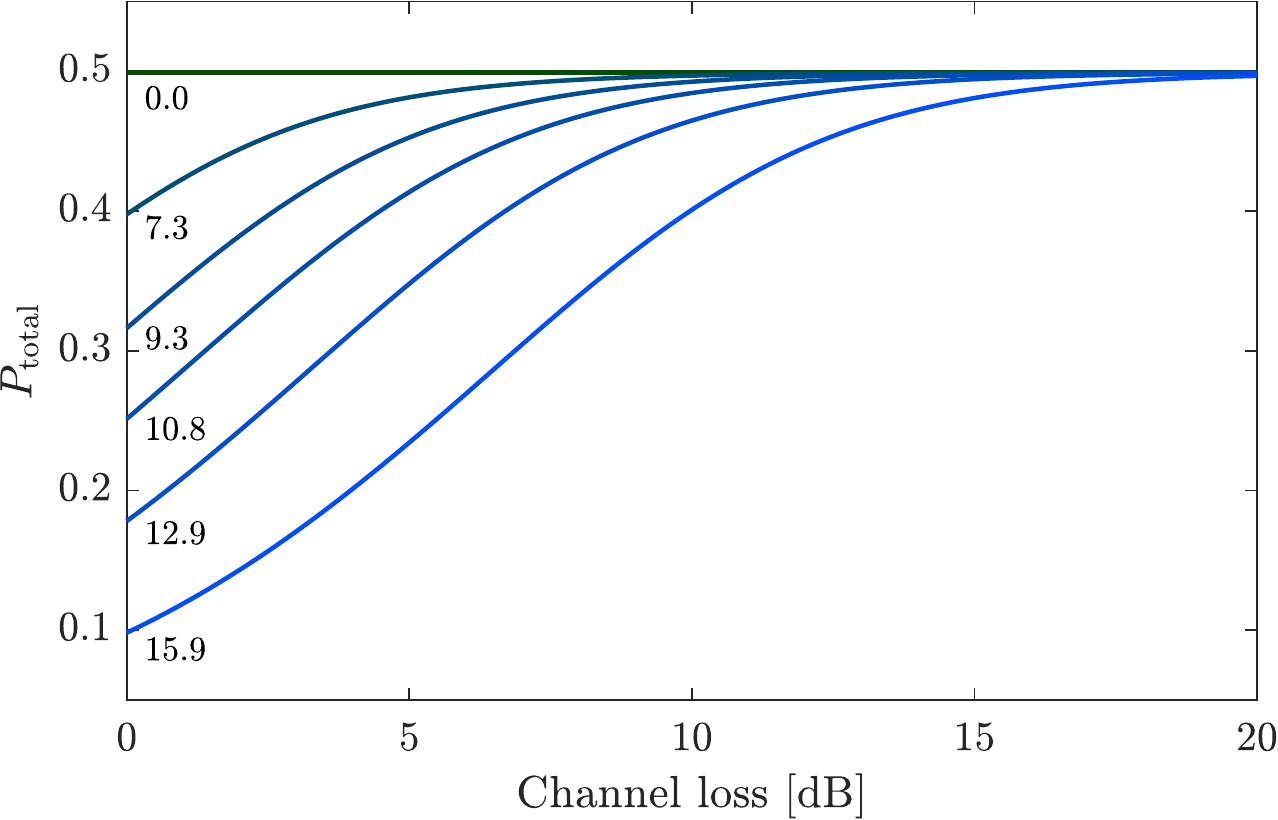}
	
	(b)
	
	\caption{(a) Average teleportation fidelity $\bar{\mathcal{F}}$ for the CV-BSM protocol (blue surface) and the H-BSM protocol (red surface). The gray plane represents the classical limit for the teleportation of qubits. 
	(b) The total success probability for the H-BSM protocol averaged over the parameters of the input qubit ($\theta$ and $\phi$).
	The numbers below each curve represents the squeezing (in dB) of the initial TMSV state.}
	\label{fig:figbellhomotmsv}
\end{figure}

Fig.~\ref{fig:figbellhomotmsv}a compares the $\bar{\mathcal{F}}$ for the CV-BSM protocol and the H-BSM protocol.
For the CV-BSM protocol, the displacement gain $g$ is set  to maximize $\bar{\mathcal{F}}$.
The CV-BSM protocol is not effective over most regions of the parameter space since it only provides a narrow region where the teleportation channel provides $\bar{\mathcal{F}}$ higher than the classical limit of $2/3$ \cite{popescu1994bell} (see Appendix B for a discussion on the classical limit).
The H-BSM protocol provides $\bar{\mathcal{F}}$ higher than $2/3$ for a region larger than the CV-BSM protocol.
Average fidelity higher than the classical limit is attainable with the H-BSM protocol when the channel loss is 4.7~dB.
The initial squeezing of the TMSV state required to provide such an $\bar{\mathcal{F}}$ is around 7~dB.

Fig.~\ref{fig:figbellhomotmsv}b shows the averaged total success probability for the H-BSM protocol against the channel loss for various values of the initial squeezing of the TMSV resource states.
The average is taken over the parameters $\theta$ and $\phi$ of the input qubit.
The H-BSM protocol provides a success probability higher than 40\% when the classical limit is surpassed.
Note, the TMSV state approaches two vacuum modes as the channel loss increases (in dB), and
the total success probability for the H-BSM protocol approaches 50\% as the multi-photon components vanish.

For the scenario of asymmetric lossy channels, the H-BSM protocol still provides $\bar{\mathcal{F}}$ higher than the classical limit for a region larger than the CV-BSM protocol.
For the  scenario where $T_1=T$ and $T_2=1$, $\bar{\mathcal{F}}$ for both protocols decreases monotonously as the channel loss increases (in dB).

\section{Teleportation with Additional Operations}\label{sec:non-Gaussian}
\subsection{The H-BSM Protocol without Entanglement Distillation}\label{sec:hbsmnoed}
\begin{figure}
	\centering
	\includegraphics[width=0.95\linewidth]{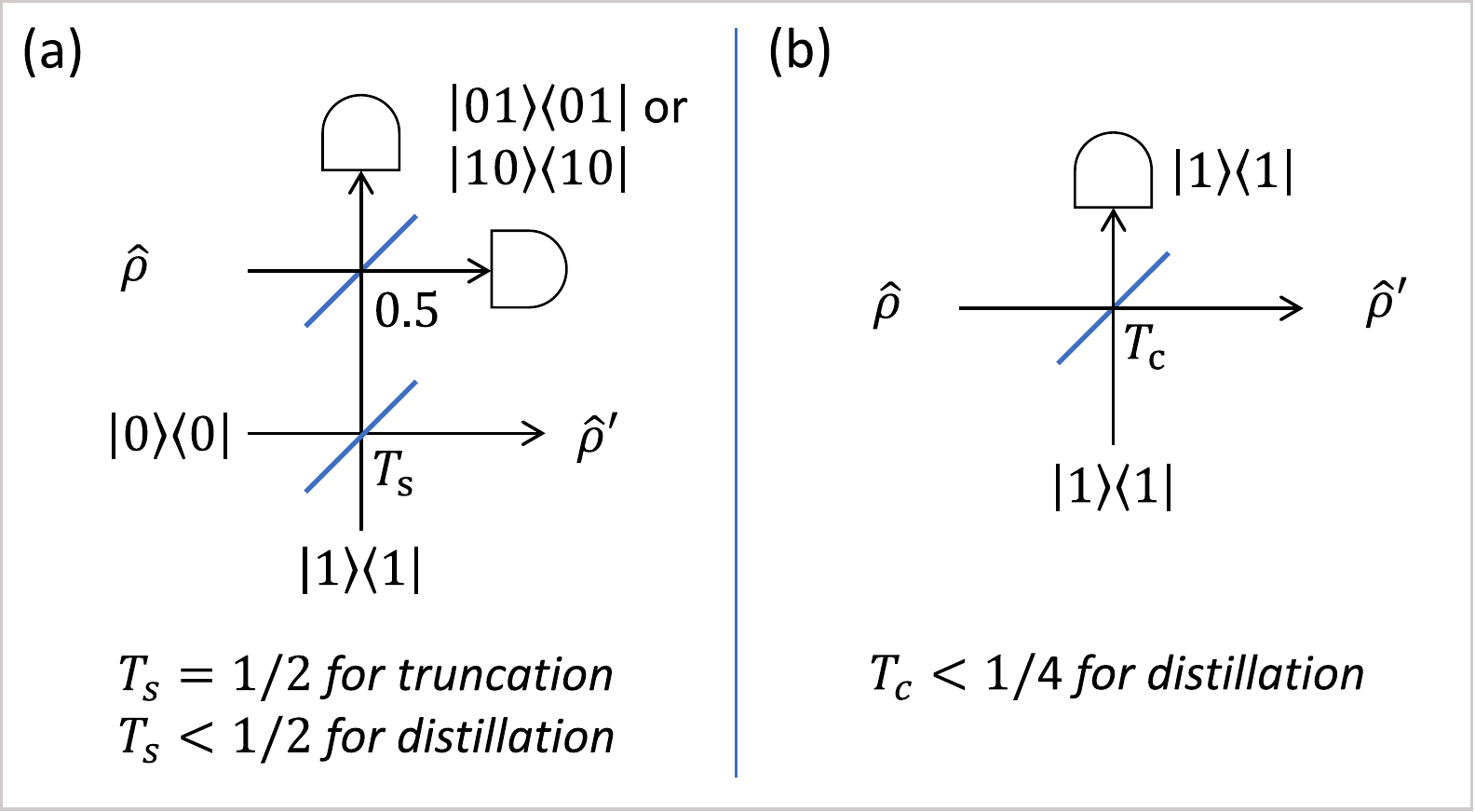}
	\caption{(a) The diagram for the QS operation on one mode of a state $\hat{\rho}$. The operation has been successful if a single photon is detected in either of the single-photon detectors.
		(b) The diagram for the PC operation. The operation has been successful if a single photon is detected in the single-photon detector. In both diagrams, the input and output to the operation are denoted by $\hat{\rho}$ and $\hat{\rho}'$, respectively.}
	\label{fig:diagqsandpc}
\end{figure}
Non-Gaussian operations are defined as operations that map a quantum state to a non-Gaussian state.
In this section, we consider the use of additional non-Gaussian operations in the H-BSM protocol, beyond the non-Gaussian photon detection utilized in the previous section.

Assume the TMSV resource state is first distributed to the two locations through lossy channels. and
similar to before, assume the two channels have the same transmissivity. In the first instance, let us also assume an H-BSM that can discriminate the four Bell states is adopted, and  that  $P_\mathrm{BSM}=1$.
A specific experimental route to the implementation of this form of the  H-BSM can be found in Appendix A.
The implementation discussed there requires the truncation of the mode of the resource channel that couples the input qubit.
The truncation can be realized by a QS operation (see Fig.~\ref{fig:diagqsandpc}a), which implements the transformation \cite{pegg1998optical}
\begin{equation}\label{eq:rhotrun}
	\hat{\rho}_{\mathrm{trun}}=\frac{1}{P_\mathrm{trun}}\hat{M}_\mathrm{trun}\hat\rho_{\mathrm{TMSV}}'\hat{M}_\mathrm{trun}^\dagger,
\end{equation}
where 
\begin{equation}
	\hat{M}_{\mathrm{trun}}=\frac{1}{\sqrt{2}}(\ket{0}_1\bra{0}+\ket{1}_1\bra{1}),
\end{equation}
$P_\mathrm{trun}=\operatorname{tr}\{(\ket{0}_1\bra{0}+\ket{1}_1\bra{1})\hat{\rho}'_{\mathrm{TMSV}}/2\}$ is the success probability for the truncation,
and $\hat\rho_{\mathrm{TMSV}}'$ is the density operator given by Eq.~(\ref{eq:rhotmsvlossy}).
The truncated TMSV state, $\hat{\rho}_{\mathrm{trun}}$, is then used as the teleportation resource channel.
The total success probability for the H-BSM protocol can then be written as $P_{\mathrm{total}}=P_{\mathrm{trun}}$.

\subsection{The H-BSM Protocol with Entanglement Distillation}
Next, consider using QS and PC to distill the entanglement of the distributed TMSV state before teleportation.
On performing the same QS operation on both modes of the distributed TMSV state, the density operator for the state can be written as \cite{ralph2009nondeterministic,hu2019entanglement}
\begin{equation}\label{eq:rhoqs}
	\hat{\rho}_{\mathrm{QS}}=\frac{1}{P_\mathrm{QS}}\hat{M}_2\hat{M}_1\hat\rho_{\mathrm{TMSV}}'\hat{M}_1^\dagger\hat{M}_2^\dagger,
\end{equation}
where $P_\mathrm{QS}$ is the overall-success probability for the two QS operations,\footnote{The overall-success probability for the two QS operations is the product of the success probabilities for each QS operation. A similar definition is used for the overall-success probability for the PC operations.}
\begin{equation}\label{eq:qsoperator}
	\hat{M}_i=\sqrt{T_\mathrm{s}}\ket{0}_k\bra{0}+\sqrt{1-T_\mathrm{s}}\ket{1}_k\bra{1}, k=1,2
\end{equation}
is the operator for the QS operation, $T_{\mathrm{s}}<1/2$ is the transmissivity of the beam-splitter used in the QS operation, and again $\hat\rho_{\mathrm{TMSV}}'$ is the density operator given by Eq.~(\ref{eq:rhotmsvlossy}).

On performing the same PC operation on both modes of the distributed TMSV state, the density operator for the resultant state can be written as \cite{hu2016multiphoton}
\begin{equation}
	\hat{\rho}_{\mathrm{PC}}=\frac{1}{P_\mathrm{PC}}\hat{R}_2\hat{R}_1\hat{\rho}_{\mathrm{TMSV}}'\hat{R}_1^\dagger\hat{R}_2^\dagger,
\end{equation}
where $P_\mathrm{PC}$ is the overall-success probability for the two PC operations,
\begin{equation}
	\hat{R}_k=\sqrt{T_\mathrm{c}}\left(\frac{T_\mathrm{c}-1}{T_\mathrm{c}}\hat{a}_k^\dagger\hat{a}_k+1\right)\sqrt{T_\mathrm{c}}^{\hat{a}_k^\dagger\hat{a}_k},
\end{equation}
is the operator for the PC operation, $T_\mathrm{c}<1/4$ is the transmissivity for the beam-splitter in PC (see Fig.~\ref{fig:diagqsandpc}b), and $\hat{a}_k^\dagger$ is the creation operator of mode $k$.

The distilled TMSV state is then used as the teleportation resource channel. 
Again we assume the form of the H-BSM that can discriminate all four Bell states is adopted.
Note, the truncation previously used is not required for a resource channel that has been distilled by QS.
In terms of total success probabilities we have the following two outcomes:
(1) $P_{\mathrm{total}}=P_\mathrm{QS}$, if the TMSV state has been distilled by QS before it is used as the resource channel. (2) $P_{\mathrm{total}}=P_\mathrm{PC}P'_\mathrm{trun}$, if the TMSV state has been distilled by PC before it is used as the resource channel, where $P'_\mathrm{trun}=\operatorname{tr}\{(\ket{0}_1\bra{0}+\ket{1}_1\bra{1})\hat{\rho}_{\mathrm{PC}}/2\}$.

\begin{figure}
	\centering
	\includegraphics[width=.95\linewidth]{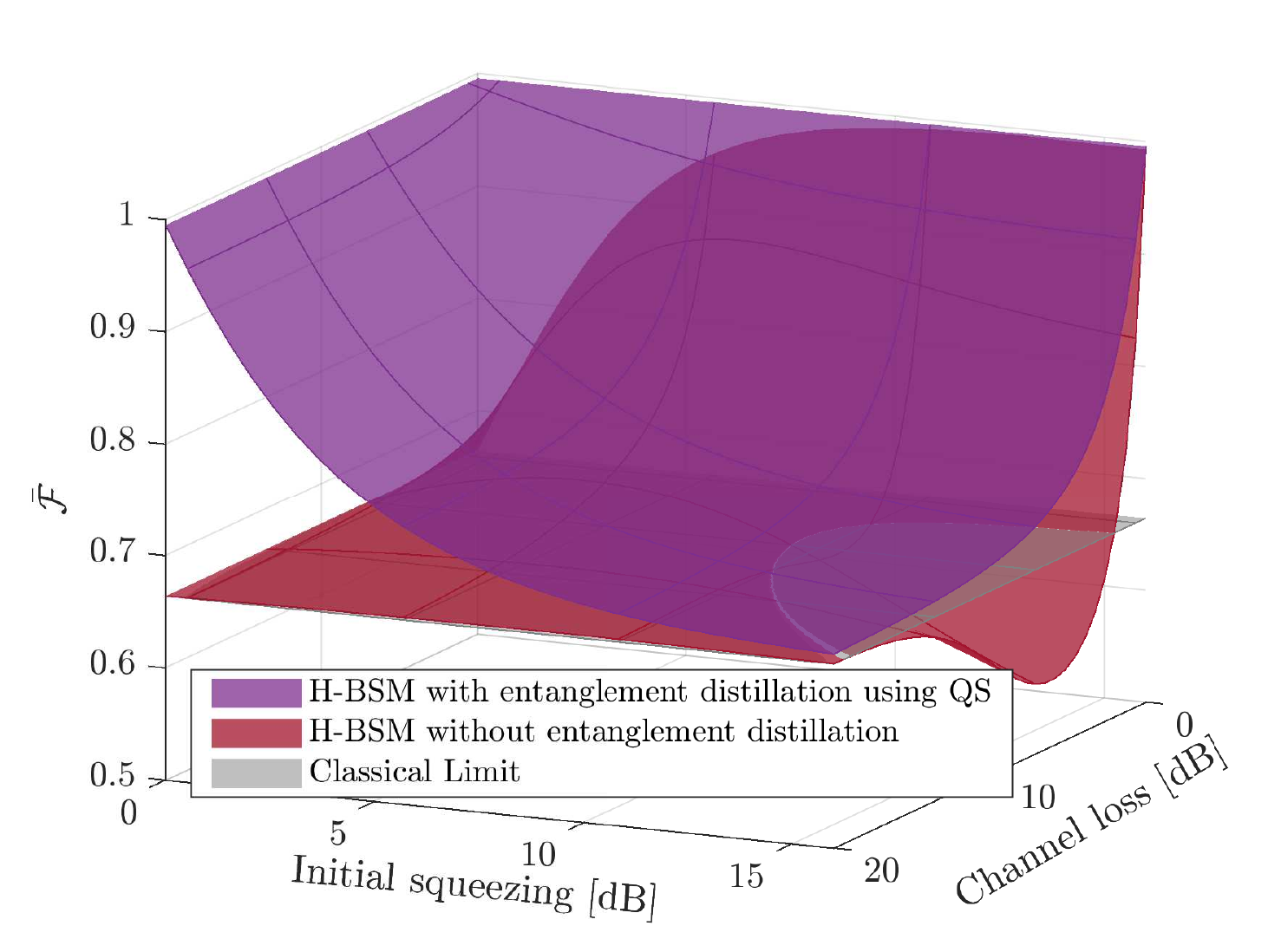}
	
	(a)
	
	\includegraphics[width=.95\linewidth]{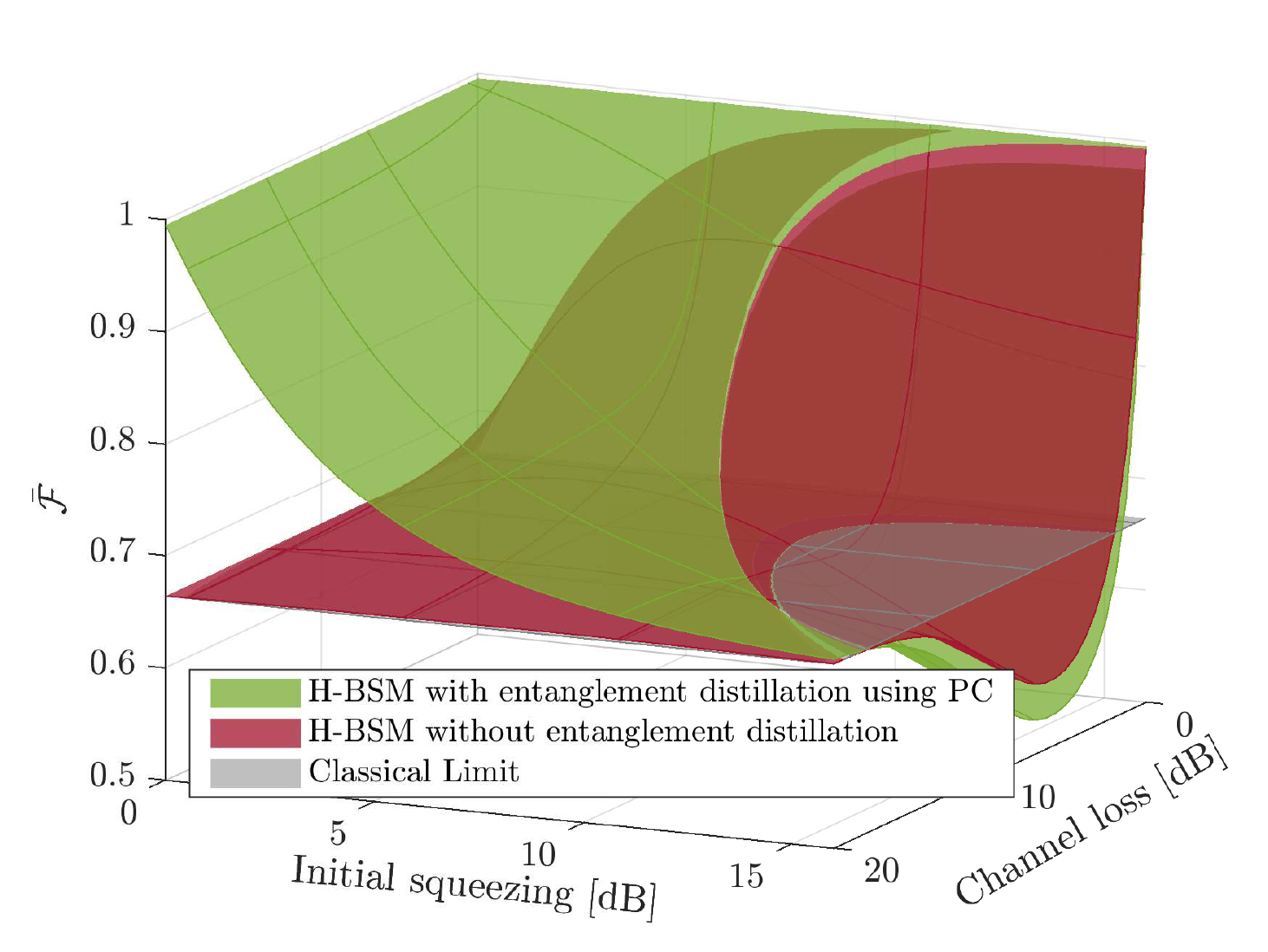}
	
	(b)
	
	\includegraphics[width=.95\linewidth]{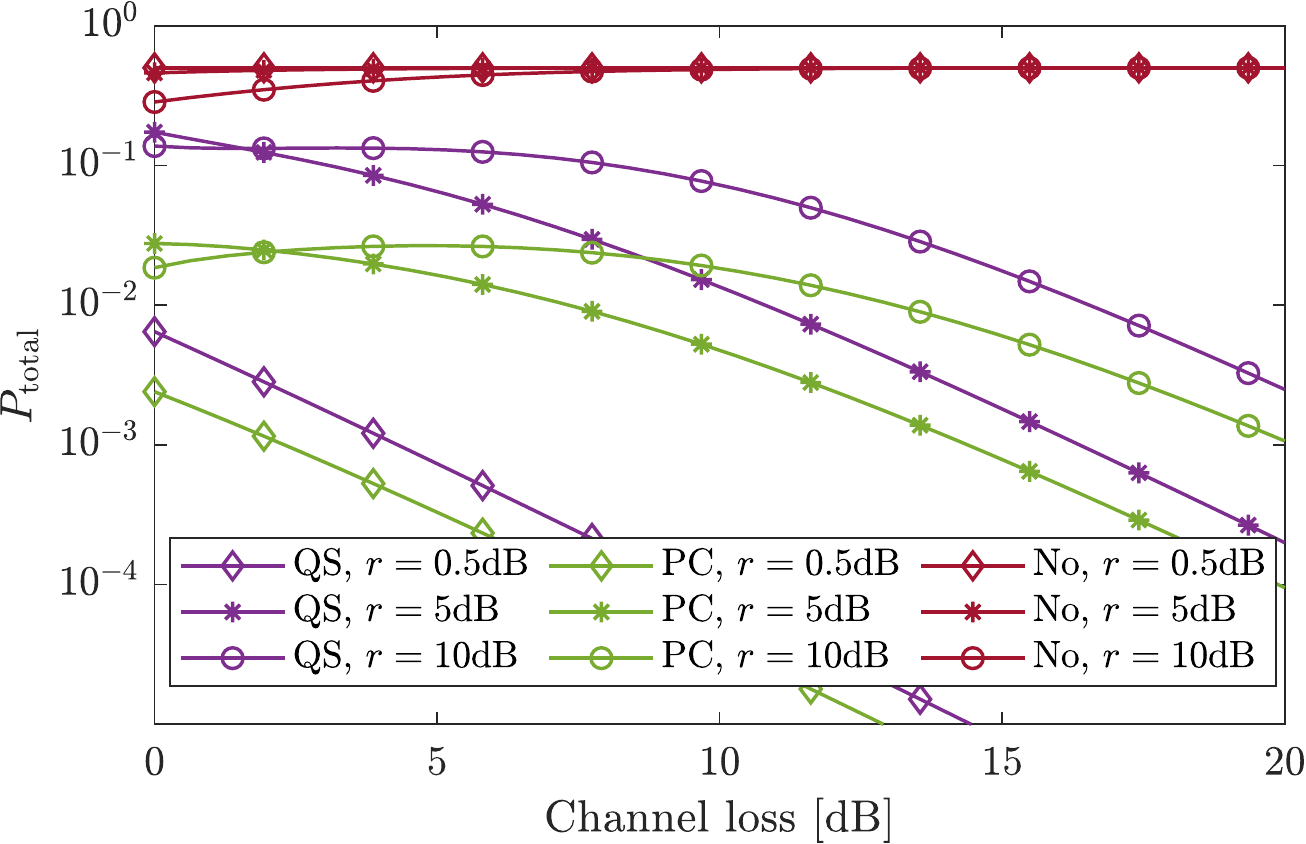}
	
	(c)
	
	\caption{(a) and (b) The average fidelity of the H-BSM protocol with QS or PC used for entanglement distillation.
		The red surface represents the  average fidelity for the case without entanglement distillation.
		The purple (green) surface represents the average fidelity for the case with QS (PC).
		The gray plane represents the classical limit.
		(c) The corresponding total success probability for the H-BSM protocol. 
		Curves with different markers represent different initial squeezing of the resource channel.
		The red curves represent the case without entanglement distillation.}
	\label{fig:bellfidelity}
\end{figure}

Fig.~\ref{fig:bellfidelity} compares the performance of the H-BSM protocol with entanglement distillation implemented by QS or PC.
In Fig.~\ref{fig:bellfidelity}a and b,
the red surface corresponds to the case without entanglement distillation.
When all four Bell states can be discriminated, the H-BSM protocol provides $\bar{\mathcal{F}}$ higher than the CV-BSM protocol over the entire region of parameter space.
Average fidelity higher than the classical limit is attainable with the H-BSM protocol even when the channel loss is 20~dB.
The purple and green surfaces correspond to the cases with QS and PC, respectively.
For QS and PC the transmissivities $T_\mathrm{s}$ and $T_\mathrm{c}$ are set  to maximize $\bar{\mathcal{F}}$.
In comparison to not performing any distillation before teleportation, both QS and PC can improve the average fidelity over some parameter spaces.
QS provides the largest $\bar{\mathcal{F}}$ over the entire parameter space. 
The average fidelity provided by QS is always larger than the classical limit.
The total success probability for the H-BSM protocol with additional non-Gaussian operations is shown in Fig.~\ref{fig:bellfidelity}c.
When the initial squeezing of the TMSV resource state approaches zero (curves marked with diamonds),
both QS and PC provide for an average fidelity approaching unity - at the cost of vanishing success probability for the operations.
The total success probability for the protocol is reduced significantly when PC is applied.

For the scenario of asymmetric lossy channels, parameters for the non-Gaussian operations performed on the two modes of the resource channel can be set independently to maximize $\bar{\mathcal{F}}$.
QS provides the largest $\bar{\mathcal{F}}$ over the entire parameter space of $\{r,T_1,T_2\}$. 
For both QS and PC, their $P_\mathrm{total}$ for the case of $T_1>T,\ T_2=T$ is always higher than the case of $T_1=T_2=T$.

\subsection{The CV-BSM Protocol}
For completeness, the use of non-Gaussian operations in the CV-BSM teleportation protocol is also studied.
To  determine fidelities, the characteristic functions for the non-Gaussified resource states are useful.
The characteristic function for the resource state after the symmetric QS operations can be directly obtained from
\begin{equation}\label{eq:qsd_tmsv}
	\bigchi_{\mathrm{QS}}(\xi_1, \xi_2)=\mathrm{tr}\left\{\hat{D}(\xi_1)\hat{D}(\xi_2)\hat{\rho}_{\mathrm{QS}}\right\},
\end{equation}
where $\hat{D}(\xi)$ is the displacement operator, and $\hat{\rho}_{\mathrm{QS}}$ is the density operator given by Eq.~(\ref{eq:rhoqs}).

Finding the characteristic function for the resultant state after the symmetric PC operations is more involved \cite{hu2020continuous} - it is given by
\begin{equation}\label{eq:pcd_tmsv}
	\begin{aligned}		
		\bigchi_{\mathrm{PC}}(\xi_1, \xi_2)
		=&\int\frac{d^2 \gamma_1 d^2\gamma_2}{\pi^2 (1-T_\mathrm c)^2}\bigchi_{\mathrm{TMSV}}'(\gamma_1, \gamma_2)\\
		&\times k(\xi_1, \gamma_1)k(\gamma_1, \xi_1)k(\xi_2, \gamma_2)k(\gamma_2, \xi_2),
	\end{aligned}
\end{equation}
where $\bigchi_{\mathrm{TMSV}}'(\gamma_1, \gamma_2)$ is the characteristic function given by Eq.~(\ref{eq:tmsv_lossy}), $T_\mathrm{c}$ is the transmissivity for the beam-splitter in PC, and
\begin{equation}
	k(\xi,\gamma)=\frac{\xi-\gamma\sqrt{T_\mathrm{c}}}{\sqrt{1-T_\mathrm{c}}}.
\end{equation}
For a given resource channel, depending on which non-Gaussian operations is performed, the characteristic function for the distilled resource channel can be obtained by putting Eq.~(\ref{eq:qsd_tmsv}) or Eq.~(\ref{eq:pcd_tmsv}) into Eq.~(\ref{eq:cfinandout}).

\begin{figure}
	\centering
	\includegraphics[width=.95\linewidth]{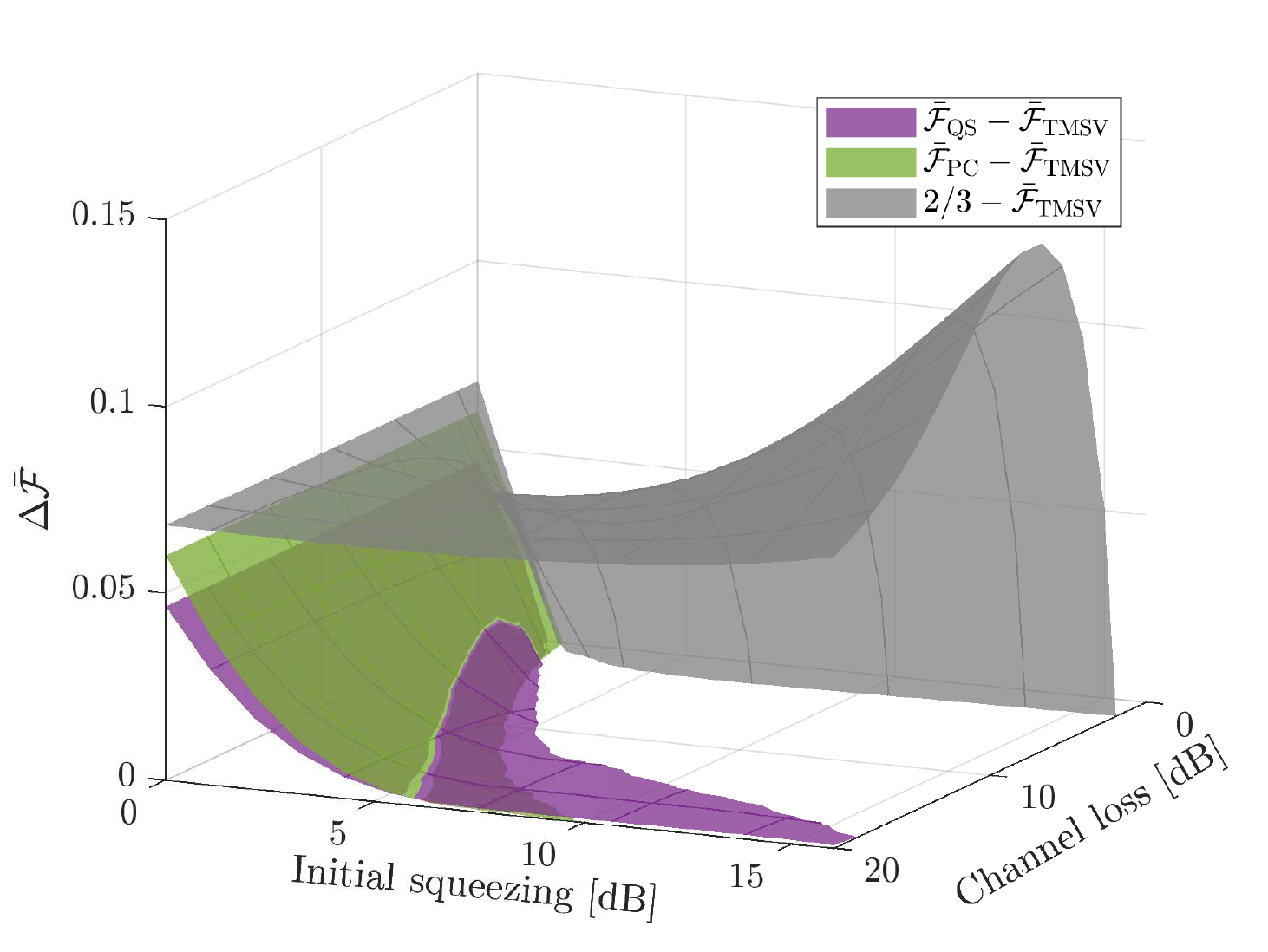}
	\caption{The difference of the average fidelity of the CV-BSM protocol with QS or PC. 
		The gray surface represents the difference between the classical limit $2/3$ and the average fidelity for the case without entanglement distillation.
		The purple (green) surface represents the difference between the average fidelity for the case with QS (PC) and the case without entanglement distillation.
		}
	\label{fig:homofidelity}
\end{figure}

Fig.~\ref{fig:homofidelity} compares the performance of the CV-BSM protocol with QS or PC.
Different from Fig.~\ref{fig:bellfidelity}, for added clarity, the difference between the average fidelity for the cases with QS or PC and the average fidelity for the case without entanglement distillation is shown.
Both QS and PC can improve the average fidelity over a certain range of the parameter space.
The improvement is the most significant when the initial squeezing of the TMSV state approaches zero. 
However, at such squeezing this improvement is meaningless as none of the non-Gaussian operations can provide average fidelity larger than the classical limit.

\section{The Impact of Imperfect Single-Photon Detectors}\label{sec:ineffdetector}
\begin{figure}
	\centering
	\includegraphics[width=.95\linewidth]{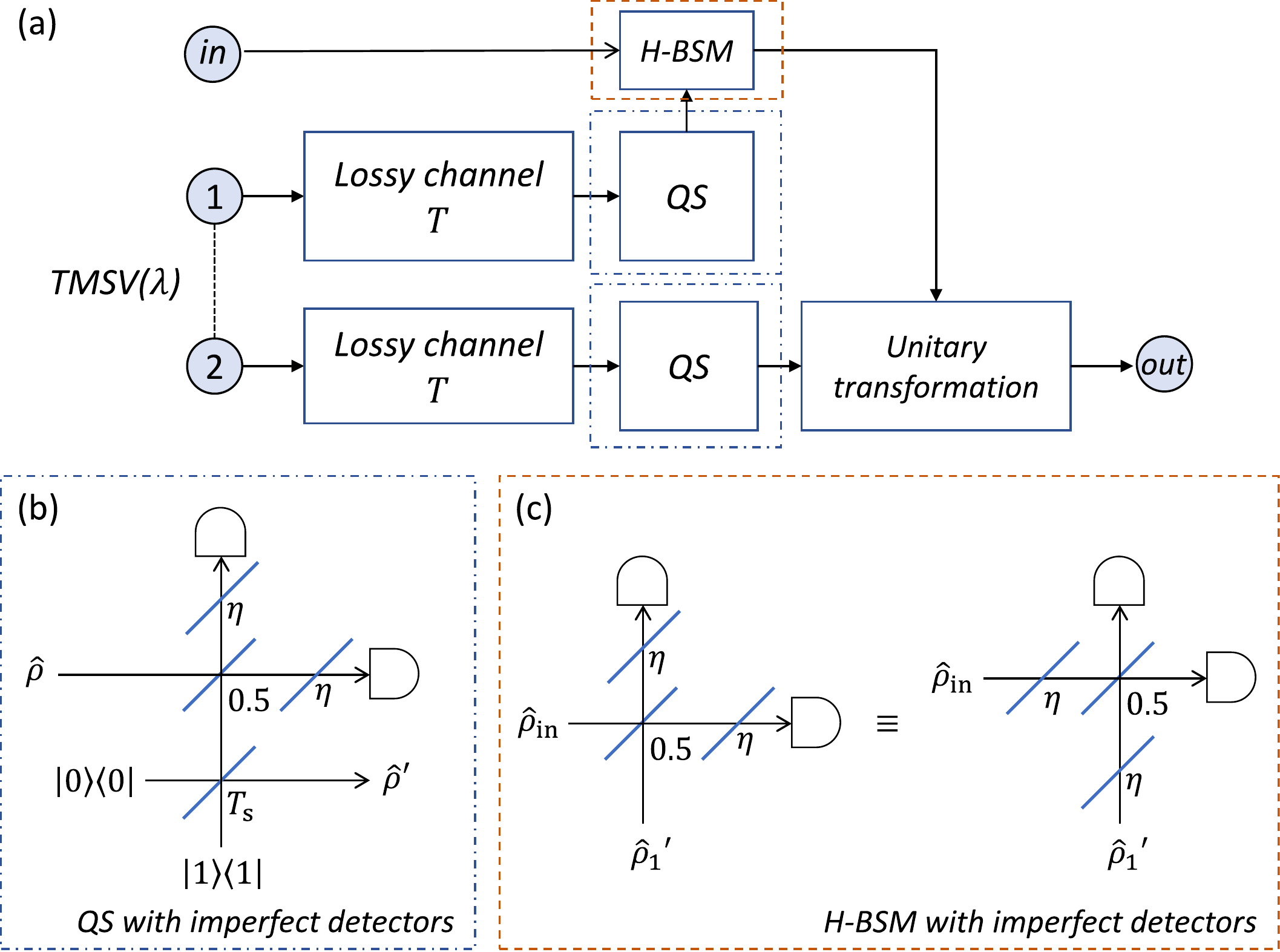}
	\caption{(a) The H-BSM protocol with QS operations and the H-BSM that only discriminates the Bell states $\ket{\Psi^\pm}$. 
	(b) The QS operation with imperfect single-photon detectors.
	(c) The H-BSM that only discriminates the Bell states $\ket{\Psi^\pm}$ with imperfect single-photon detectors.
		}
	\label{fig:tp_efficiency}
\end{figure}

Consider the use of single-photon detectors with detection efficiency $\eta$.\footnote{The detection efficiency of the single-photon detector, which is defined as the probability of registering a count if a photon arrives at the detector, is considered as the product of coupling losses and the intrinsic quantum efficiency of the detector \cite{hadfield2009single}.} To bring focus to this discussion, as shown in Fig.~\ref{fig:tp_efficiency}a, we will consider the photon detectors used in the QS operation and the H-BSM (recall from the previous section, QS outperforms PC in terms of fidelity and success probability).
Considering that an imperfect single-photon detector, i.e., a detector with $\eta<1$, is equivalent to a detector with unity efficiency preceded by a beam-splitter with transmissivity $\eta$ \cite{hogg2014efficiencies}, we will use such an equivalent form and model the detection efficiency by an additional beam-splitter with transmissivity $\eta$.
A mode to be detected first interacts with a vacuum mode at the beam-splitter.

Note, the impact the detection efficiency has on the H-BSM depends on the implementation of the measurement \cite{wein2016efficiency}.
For simplicity, we revert in this section back to the form of the H-BSM that only discriminates the Bell states $\ket{\Psi^\pm}$.

The QS operation with imperfect single-photon detectors is shown in Fig.~\ref{fig:tp_efficiency}b. 
The teleportation resource channel after the two symmetric QS operations with the same imperfect detectors can be written as
\begin{equation}\label{eq:qsineff}
    \hat{\rho}_{\mathrm{QS}}=\mathcal{Q}_{1'}\circ\mathcal{Q}_{2'}(\hat{\rho}_{\mathrm{TMSV}}'),
\end{equation}
where the transformation for the QS operation on a mode of a state $\hat{\rho}$ can be written as \cite{he2021noiseless}
\begin{equation}\label{eq:QSineff}
	\begin{aligned}
		\hat{\rho}'&=\mathcal{Q}(\hat{\rho})\\
		&:=\frac{1}{P_{\eta}}
		\sum_{n=1,\ n'=0}^{\infty}
		n\eta
		(1-\eta)^{n'+n-1}
		\hat{M}_{n,n'}
		\hat{\rho}
		\hat{M}_{n,n'}^\dagger,
	\end{aligned}	
\end{equation}
where $P_{\eta}$ is a normalization constant, and
\begin{equation}\label{eq:qs_onoff}
	\begin{aligned}			
		\hat{M}_{n,n'}=&\left(-1\right)^{n'}{2}^{-\frac{n+n'-1}{2}}(n-n')\sqrt{\frac{\left(n+n'-1\right)!}{n!n'!}}\\
		&\times		
		\sqrt{T_\mathrm{s}}\ket{0}\bra{n+n'-1}\\
		&+
		\left(-1\right)^{n'}
		{2}^{-\frac{n+n'-1}{2}}\sqrt{\frac{(n+n')!}{n!n'!}}
		\sqrt{1-T_\mathrm{s}}\ket{1}\bra{n+n'},
	\end{aligned}
\end{equation}
which reduces to the operator given by Eq.~(\ref{eq:qsoperator}) when $\eta=1$.
We note, four imperfect single-photon detectors are used in the two QS operations.

As shown in Fig.~\ref{fig:tp_efficiency}c, suppose the H-BSM adopts single-photon detectors with the same detection efficiency as the QS operations.
Let $\hat{\rho}_{\mathrm{det}}$ be the full description of the system before the H-BSM with imperfect detectors,
\begin{equation}
	\hat{\rho}_{\mathrm{det}}=\hat{\rho}_\mathrm{in}\otimes\hat{\rho}_{\mathrm{QS}},
\end{equation}
where $\hat{\rho}_\mathrm{in}$ is the qubit to be teleported, and $\hat{{\rho}}_\mathrm{QS}$ is the density operator given by Eq.~(\ref{eq:qsineff}).
We note that the beam-splitters modeling the detection efficiency commute with the 50:50 beam-splitter in the H-BSM \cite{hogg2014efficiencies}.
To simplify the analysis we swap the order of the beam-splitters with transmissivity $\eta$ and the 50:50 beam-splitter.
The beam-splitters with transmissivity $\eta$ alter the state $\hat{\rho}_{\mathrm{det}}$ to
\begin{equation}
	\hat{\rho}_{\mathrm{eq}}=\mathcal{E}_\mathrm{in}\circ\mathcal{E}_{1'}(\hat{\rho}_{\mathrm{det}}),
\end{equation}
where the transformation $\mathcal{E}$ is similar to the transformation given by Eq.~(\ref{eq:lossyopeator}), but with $T_i$ being replaced by $\eta$. 
An H-BSM using efficient single-photon detectors ($\eta=1$) is then performed on mode `in' and mode $1'$ of $\hat{\rho}_{\mathrm{eq}}$.

\begin{figure}
	\centering
	\includegraphics[width=.85\linewidth]{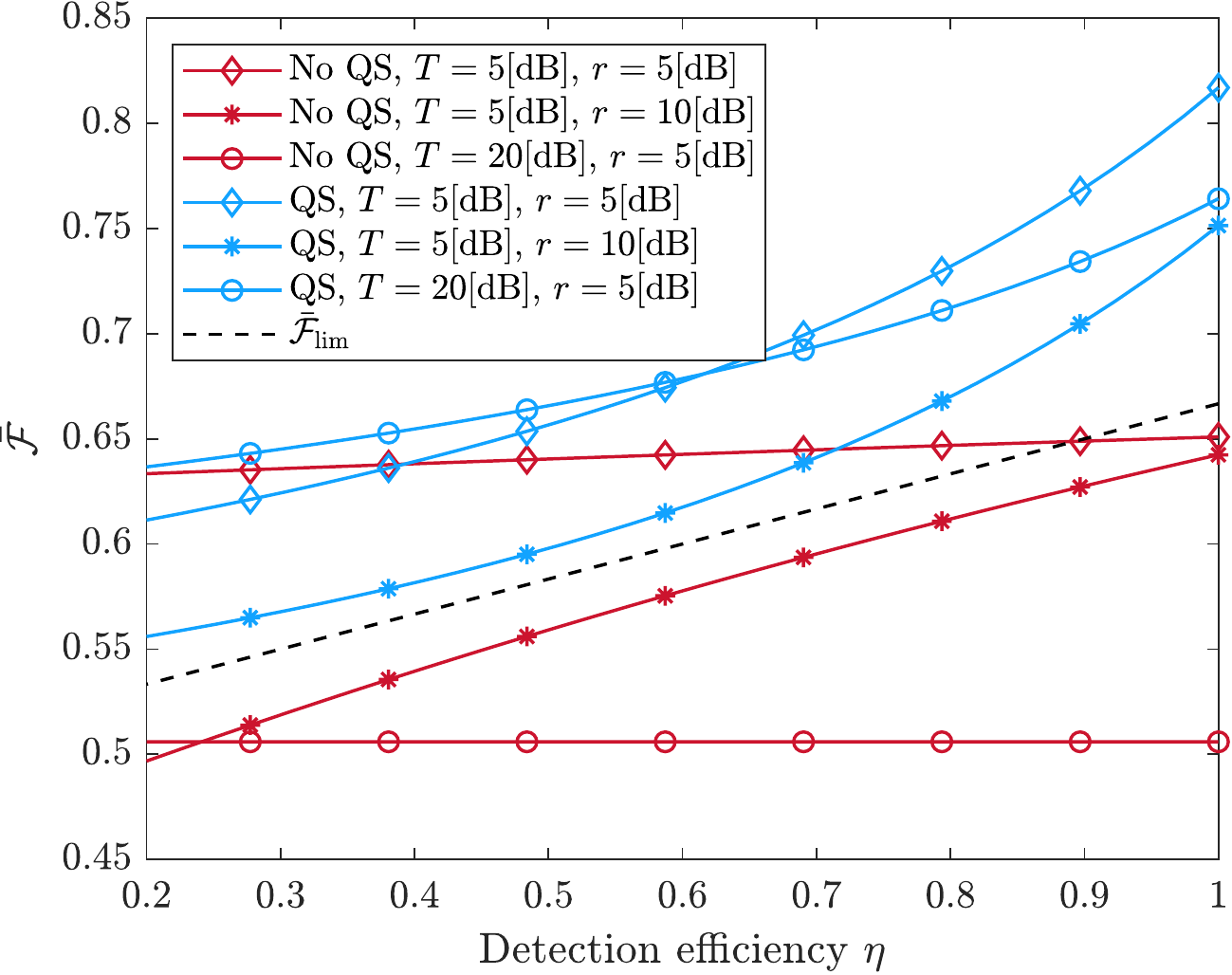}
	
	(a)
	
	\includegraphics[width=.85\linewidth]{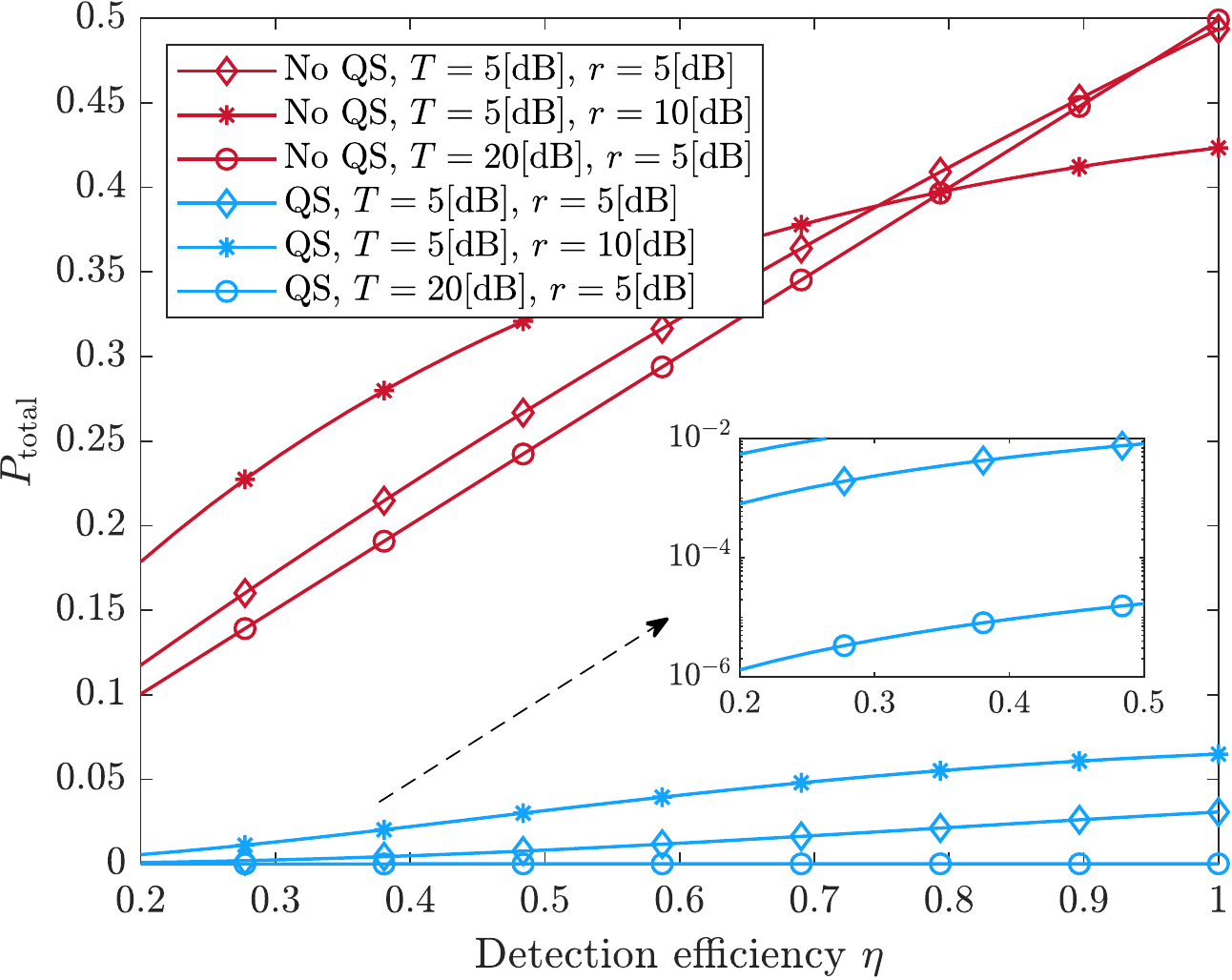}
	
	(b)
	
	\caption{(a) The average fidelity for the H-BSM protocol with imperfect single-photon detectors for different initial squeezing ($r$) and channel loss ($T$).
	The red curves represent the case of no entanglement distillation (QS) adopted.
	The blue curves represent the case with QS.
	The black dashed line represents the classical limit $\bar{\mathcal{F}}_{\mathrm{lim}}$.
	(b) The corresponding total success probability for the H-BSM protocol.}
	\label{fig:bellfidelityineff}
\end{figure}

Consider the scenario of symmetric lossy channels.
The average fidelity for the H-BSM protocol with imperfect single-photon detectors is shown in Fig.~\ref{fig:bellfidelityineff}a.
For the case of no QS adopted, 
The average fidelity is not sensitive to the decrease of $\eta$ when the initial squeezing is small.
With QS the H-BSM protocol is more robust against the decrease of $\eta$.
Average fidelity higher than the classical limit, $\bar{\mathcal{F}}_{\mathrm{lim}}$, is attainable even when $\eta$ drops to 0.2 (see Appendix B for a discussion on the classical limit).
When single-photon detectors with $\eta\approx0.9$ (e.g. \cite{marsili2013detecting}) are used, the H-BSM protocol with QS provides average fidelity higher than the classical limit over the entire channel loss region considered (up to 20dB loss, which is similar to the H-BSM protocol discussed in Section.~\ref{sec:hbsmnoed}).

Fig.~\ref{fig:bellfidelityineff}b shows the corresponding total success probability for the H-BSM protocol.
For both cases, $P_{\mathrm{total}}$ decreases monotonously as $\eta$ decreases, independent of the channel loss level and the initial squeezing.

\section{Conclusion}\label{sec:conclusion}
We have proposed a new hybrid teleportation protocol for DV qubits that combines a CV teleportation resource channel and a  new form of a hybrid Bell state measurement. We discussed two forms of this measurement, one in which only two of the Bell states are discriminated, and a more complex one in which all Bell states are discriminated.
We found that the loss tolerance in our protocol was significantly improved relative to known teleportation protocols and that high fidelity and success probability were both possible.	
The use of  non-Gaussian operations on the resource channel was also studied. 
With such operations, a teleportation fidelity approaching unity was found possible at any channel loss level, albeit at vanishing success probability.
Finally, we studied the impact of single-photon detection with non-unity efficiency in our protocol, finding that it can accommodate significant channel losses using current single-photon detectors.

\appendix
\section{Implementation of the H-BSM}\label{apx:dvbsmimplementation}
Here, we discuss the implementation of the H-BSM. First, consider the form of the H-BSM that can only discriminate the Bell States $\ket{\Psi^\pm}$ \cite{lombardi2002teleportation}, which contains a beam-splitter and two single-photon detectors.
We define $n_i$ as the number of photons detected at the $i$-th detector, $\hat{a}_1^\dagger$ as the creation operator for the qubit to be teleported, and $\hat{a}_2^\dagger$ as the creation operator for the mode of the teleportation resource channel that couples the qubit.
Modes $1$ and $2$ are mixed at the beam-splitter, which implements the transformation,
\begin{equation}
    \begin{pmatrix}
        \hat{b}_1^\dagger\\\hat{b}_2^\dagger
    \end{pmatrix}=
    \mathbf{S}^{(1)}
    \begin{pmatrix}
        \hat{a}_1^\dagger\\\hat{a}_2^\dagger
    \end{pmatrix},
\end{equation}
where
\begin{equation}\mathbf{S}^{(1)}=\frac{1}{\sqrt{2}}\begin{pmatrix}
	1&1\\-1&1
\end{pmatrix},
\end{equation}
and $\hat{b}_1^\dagger$ and $\hat{b}_2^\dagger$ are the creation operators for the output modes of the beam-splitter.

The  H-BSM is successful when one photon is detected by one of the detectors.
For $n_1=1$ and $n_2=0$, the input modes are projected into
\begin{equation}
    \hat{b}_1^\dagger\ket{00}=\frac{1}{\sqrt{2}}\left(\hat{a}_1^\dagger+\hat{a}_2^\dagger\right)\ket{00}=\frac{1}{\sqrt{2}}\left(\ket{10}+\ket{01}\right).
\end{equation}
For $n_1=0$ and $n_2=1$, the input modes are projected into $\frac{1}{\sqrt2}\left(\ket{10}-\ket{01}\right)$.
The probabilities for the projection of $\ket{\Psi^+}$ and $\ket{\Psi^-}$ are, respectively,
\begin{equation}
	\begin{aligned}
		P_{\Psi^+}=\operatorname{tr}\{\ket{\Psi^+}\bra{\Psi^+}\hat{\rho}_{12}\},\\
		P_{\Psi^-}=\operatorname{tr}\{\ket{\Psi^-}\bra{\Psi^-}\hat{\rho}_{12}\},
	\end{aligned}
\end{equation}
where $\hat{\rho}_{12}$ is the density operator for modes 1 and 2. 
When no other non-Gaussian operations are adopted, the total success probability for the H-BSM protocol can then be written as $P_{\mathrm{total}}=P_{\Psi^+}+P_{\Psi^-}$.

\begin{figure}
	\centering
	\includegraphics[width=0.95\linewidth]{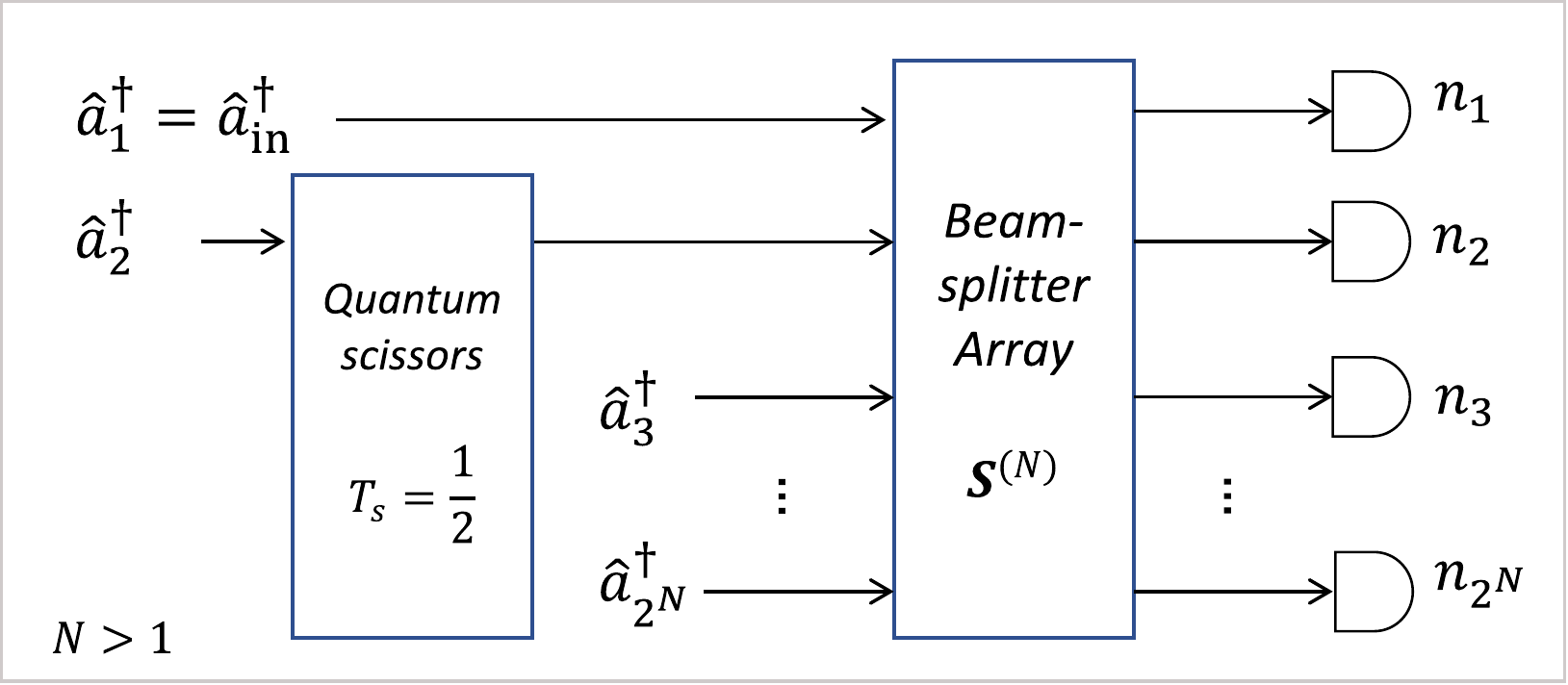}
	\caption{The diagram for the H-BSM, where  $\hat{a}_1^\dagger$ and $\hat{a}_2^\dagger$ are the creation operators  for the qubit to be teleported and the mode of the resource channel that couples the qubit, respectively, 
	$\hat{a}_3^\dagger,\hat{a}_4^\dagger,\dots,\hat{a}_{2^N}^\dagger$ are the creation operators for the ancillary modes, $T_\mathrm{s}$ is a parameter associated with the QS, 
	and $n_i$ is the photon number detected at the $i$-th single-photon detector. 
	The mode of the resource channel is first truncated by a QS operation before entering the beam-splitter array.}
	\label{fig:diagdvbsm}
\end{figure}

Next, built upon \cite{grice2011arbitrarily}, we present a variation of the H-BSM that can discriminate the four Bell states expressed in the Fock basis (see also \cite{olivo2018ancilla}). We note that this variation is only one of the many routes possible to such a full-discrimination measurement.
As shown in Fig.~\ref{fig:diagdvbsm}, this H-BSM consists of an array of beam-splitters ($2^{2N-2}$ in total), $2^N-2$ ancillary modes, and $2^N$ single-photon detectors ($N>1$).
Same as before, the creation operator for the qubit to be teleported is denoted by $\hat{a}_1^\dagger$ and the creation operator for the mode of the teleportation resource channel that couples the qubit is denoted by $\hat{a}_2^\dagger$.
The creation operators for the ancillary modes are denoted by $\hat{a}_3^\dagger,\hat{a}_4^\dagger,\dots,\hat{a}_{2^N}^\dagger$.
Mode $2$ should be truncated to the space spanned by $\{\ket{0},\ket{1}\}$ before it enters the beam-splitter array.
This can be implemented by performing a QS operation with $T_\mathrm{s}=0.5$ on mode $2$.
Let $\hat{\rho}_{2}$ be the density operator of mode $2$.
The success probability for such a QS operation can then be written as
\begin{equation}
    P_\mathrm{trun}=\frac{1}{2}\operatorname{tr}\{\left(\ket{0}\bra{0}+\ket{1}\bra{1}\right)\hat{\rho}_{2}\}.
\end{equation}

The beam-splitter array implements the transformation,
\begin{equation}
	\begin{pmatrix}
		\hat{b}_1^\dagger\\\hat{b}_2^\dagger\\\vdots\\\hat{b}_{2^N}
	\end{pmatrix}
	=\mathbf{S}^{(N)}
	\begin{pmatrix}
		\hat{a}_1^\dagger\\\hat{a}_2^\dagger\\\vdots\\\hat{a}_{2^N}
	\end{pmatrix},
\end{equation}
where 
\begin{equation}
	\mathbf{S}^{(N)}=\mathbf{S}^{(1)}\otimes\mathbf{S}^{(N-1)},
\end{equation}
and $\hat{b}_1^\dagger,\hat{b}_2^\dagger,\dots,\hat{b}_{2^N}^\dagger$ are the creation operators for the output modes of the beam-splitter array.

Before entering the beam-splitter array, the $2^N-2$ ancillary modes are in the state
\begin{equation}
	\ket{\Xi^{(N)}}=\bigotimes_{j=1}^{N-1}\ket{\xi^{(j)}},
\end{equation}	
which is a product state of $N-1$ different entangled states,
\begin{equation}
	\ket{\xi^{(j)}}=
	\frac{1}{\sqrt{2}}\left(
	\underbrace{\ket{00\dots0}}_{2^{j}}+ 
	\underbrace{\ket{11\dots1}}_{2^{j}}
	\right).
\end{equation}

We label the summation of the numbers of photons detected by the detectors as $n_\mathrm{sum}=\sum_{i=1}^{2^N} n_i\le 2^N$.
On the detection of $\bigotimes_{i=1}^{2^N}\ket{n_i}$ at the output of the beam-splitter array,
modes $1$ and $2$ are projected into the state
\begin{equation}\label{eq:dvbsmp}
	\hat{\rho}_{12}'=\frac{\hat{\Pi}\hat{\rho}_{12}\hat{\Pi}^\dagger}{\operatorname{tr}\{\hat{\Pi}\hat{\rho}_{12}\}},
\end{equation}
where $\hat\Pi=\ket{\zeta}\bra{\zeta}$, and
\begin{equation}
	\ket{\zeta}=
	\bra{\Xi^{(N)}}
	\frac{1}{\sqrt{2}^{n_\mathrm{sum}}}
	\prod_{i=1}^{2^N}\frac{\left(\sum_{j=1}^{2^N}s^{(N)}_{i,j}\hat{a}_j^\dagger\right)^{n_i}}{\sqrt{n_i!}}\ket{0},
\end{equation}
where $s^{(N)}_{i,j}$ are elements of the matrix $\mathbf{S}^{(N)}$.

Depending on the value of $n_\mathrm{sum}$, the detection outcomes can be categorized into four mutually exclusive groups:

\noindent
(1) When $n_\mathrm{sum}=0$, modes 1 and 2 are projected into $\ket{00}$. 
The probability for this group is
\begin{equation}
P_{00}=\frac{\operatorname{tr}\{\ket{00}\bra{00}\hat{\rho}_{12}\}}{2^{N-1}}.
\end{equation} 

\noindent
(2) When $n_\mathrm{sum}=2^N$, independent of the permutation of $\{n_i\}$, modes 1 and 2 are projected into $\ket{11}$.
The probability for this group is
\begin{equation}
	P_{11}=\frac{\operatorname{tr}\{\ket{11}\bra{11}\hat{\rho}_{12}\}}{2^{N-1}}.
\end{equation} 

\noindent
(3) When $n_\mathrm{sum}$ is even (but not equal to $0$ and $2^N$), depending on the permutation of $\{n_i\}$, modes 1 and 2 are projected into $\ket{\Phi^+}$ or $\ket{\Phi^-}$.
For a given permutation of $\{n_i\}$, the states $\ket{\Phi^+}$ and $\ket{\Phi^-}$ can be determined by using Eq.~(\ref{eq:dvbsmp}).
The probabilities for the projection of $\ket{\Phi^+}$ and $\ket{\Phi^-}$ are, respectively,
\begin{equation}
	\begin{aligned}		
	P_{\Phi^+}=\left(1-\frac{1}{2^{N-1}}\right)\operatorname{tr}\{\ket{\Phi^+}\bra{\Phi^+}\hat{\rho}_{12}\},\\		P_{\Phi^-}=\left(1-\frac{1}{2^{N-1}}\right)\operatorname{tr}\{\ket{\Phi^-}\bra{\Phi^-}\hat{\rho}_{12}\}.		
	\end{aligned}
\end{equation}

\noindent
(4) When $n_\mathrm{sum}$ is odd, depending on the permutation of $\{n_i\}$, modes 1 and 2 are projected into $\ket{\Psi^+}$ or $\ket{\Psi^-}$.
For a given permutation of $\{n_i\}$, the states $\ket{\Psi^+}$ and $\ket{\Psi^-}$ can be determined by using Eq.~(\ref{eq:dvbsmp}).
The probabilities for the projection of $\ket{\Psi^+}$ and $\ket{\Psi^-}$ are, respectively,
\begin{equation}
	\begin{aligned}
		P_{\Psi^+}=\operatorname{tr}\{\ket{\Psi^+}\bra{\Psi^+}\hat{\rho}_{12}\},\\
		P_{\Psi^-}=\operatorname{tr}\{\ket{\Psi^-}\bra{\Psi^-}\hat{\rho}_{12}\}.		
	\end{aligned}
\end{equation}

The H-BSM is successful when modes 1 and 2 are projected into one of the four Bell states.
The corresponding success probability can be written as
\begin{equation}
	P_{\mathrm{BSM}}=P_{\Phi^+}+P_{\Phi^-}+P_{\Psi^+}+P_{\Psi^-}=1-\frac{P_\mathrm{id}}{2^{N-1}},
\end{equation}
where $P_\mathrm{id}=\operatorname{tr}\{\left(\ket{00}\bra{00}+\ket{11}\bra{11}\right)\hat{\rho}_{12}\}$.
When no other non-Gaussian operations are adopted, the total success probability for the H-BSM protocol can then be written as $P_{\mathrm{total}}=P_{\mathrm{trun}}P_{\mathrm{BSM}}$, which approaches $P_{\mathrm{trun}}$ as $N$ increases.

\section{The Classical Limit for the Teleportation of Qubits}\label{apx:limit}
Consider a qubit expressed in the Fock basis as $\ket{\mathrm{in}}=\cos(\theta/2)\ket{0}+\exp{(i\phi)}\sin(\theta/2)\ket{1}$.
Alice is given $\ket{\mathrm{in}}$, unknown to her, and Bob is required to prepare a qubit $\ket{\mathrm{out}}$ as close as possible to $\ket{\mathrm{in}}$.
Alice and Bob are connected only by a classical channel and their goal is to maximize the fidelity $|\braket{\mathrm{out}|\mathrm{in}}|^2$ averaged over the parameters $\theta$ and $\phi$.
We denote the classical limit, which is the upper limit for the average fidelity, as $\bar{\mathcal{F}}_\mathrm{lim}$.
One strategy for reaching $\bar{\mathcal{F}}_\mathrm{lim}$ is as follows \cite{popescu1994bell}.
Alice measures $\ket{\mathrm{in}}$ by a single-photon detector, of which the result can be $\ket{0}$ or $\ket{1}$.
Alice informs Bob about her measurement result, and Bob prepares a $\ket{0}$ (or $\ket{1}$) if the result is $\ket{0}$ (or $\ket{1}$).

Now consider the use of a single-photon detector with non-unity efficiency.
Again, we use a beam-splitter with transmissivity $\eta$ to model the efficiency in the detector.
For a given $\ket{\mathrm{in}}$, using such a detector the probabilities for the detection of $\ket{0}$ and $\ket{1}$ are, respectively,
\begin{equation}
    \begin{aligned}
        P_0&=\operatorname{tr}\{\ket{0}\bra{0}\mathcal{E}(\ket{\mathrm{in}}\bra{\mathrm{in}})\}=\cos^2\frac{\theta}{2}+(1-\eta)\sin^2\frac{\theta}{2},\\
        P_1&=\operatorname{tr}\{\ket{1}\bra{1}\mathcal{E}(\ket{\mathrm{in}}\bra{\mathrm{in}})\}=\eta\sin^2\frac{\theta}{2},
    \end{aligned}
\end{equation}
where again the transformation $\mathcal{E}$ is similar to the transformation given by Eq.~(\ref{eq:lossyopeator}), but with $T_i$ being replaced by $\eta$. 
The classical limit for the average fidelity can then be written as
\begin{equation}
    \begin{aligned}
        \bar{\mathcal{F}}_{\mathrm{lim}}=&
        \int_0^\pi d\theta \int_0^{2\pi} d\phi
        \left(\frac{\sin\theta}{4\pi}\right)
        \left(\cos^2\frac{\theta}{2} P_0 + \sin^2\frac{\theta}{2} P_1\right)\\
        =&\frac{3+\eta}{6},
    \end{aligned}
\end{equation}
The classical limit is $2/3$ when an efficient single-photon detector ($\eta=1$) is used.
When $\eta=0$, Bob will prepare a qubit in the state of $\ket{0}$ independent of the state of Alice's initial qubit $\ket{\mathrm{in}}$.
The corresponding classical limit is $1/2$.

\bibliography{mybib}

\end{document}